\documentclass[12pt]{iopart}

\usepackage{epsfig}
\usepackage[latin1]{inputenc}
\usepackage{amsfonts}
\usepackage{amssymb}
\usepackage{calrsfs}
\newcommand{\beq}{\begin{equation}}
\newcommand{\eeq}{\end{equation}}
\newcommand{\kB}{k_{\mbox{\tiny B}}}
\newcommand{\la}{\langle}
\newcommand{\ra}{\rangle}

\begin{document}
\title[Complex heat capacity and entropy production of temperature modulated systems]
{Complex heat capacity and entropy production of temperature modulated systems}

\author{M\'{a}rio J. de Oliveira}

\address{Instituto de F\'{\i}sica,
Universidade de S\~{a}o Paulo, \\
Rua do Mat\~ao, 1371,
05508-090 S\~{a}o Paulo, S\~{a}o Paulo, Brazil}

\ead{oliveira@if.usp.br}

\begin{abstract}

Non-equilibrium systems under temperature modulation are investigated
in the light of the stochastic thermodynamics. We show that, for small
amplitudes of the temperature oscillations, the heat flux 
behaves sinusoidally with time, a result that allows the definition of the
complex heat capacity. The real part of the complex heat capacity
is the dynamic heat capacity,
and its imaginary part is shown to be proportional to
the rate of entropy production. We also show that
the poles of the complex heat capacity are equal to imaginary unit
multiplied by the eigenvalues of the unperturbed evolution
operator, and are all located in the lower half plane of the complex
frequency, assuring that the Kramers-Kronig relations are obeyed.
We have also carried out an exact calculation of the complex heat
capacity of a harmonic solid and determined the dispersion 
relation of the dynamic heat capacity and of the rate of entropy production.

\end{abstract}

\maketitle

\section{Introduction}

Among the various methods used in calorimetry, there is one in which
the heat flow and temperature are modulated in time, oscillating around
their mean values.
Modulation calorimetry \cite{gmelin1997,kraftmakher2004}
was introduced by Corbino 
and consisted of measuring the temperature modulation
through the oscillations of the electrical resistance 
in metal filaments fed by an alternating electrical current.
By this method he measured the specific heat of tungsten
as did others later on by similar methods
\cite{worthing1918,gaehr1918,smith1922,bockstahler1925,zwikker1928}.
Other methods of modulated calorimetry were developed
\cite{filippov1966,sullivan1968}, including the so-called
temperature-modulated differential scanning calorimetry
\cite{gill1993,hohne2003,gobrecht1971}.

The response of a system to a thermal perturbation is represented
by the heat capacity $C$. As the amount of heat exchanged
and the temperature may not vary slowly on time, $C$ is defined
as the ratio between the heat flux and the time derivative of
the temperature. It is called {\it dynamic} heat capacity and should be
distinguished from the equilibrium or static heat capacity $C_0$.
Nevertheless, it approaches the static heat capacity when the
variation in temperature is slow enough. 

The time variation of $C$ may be so fast on the experimental time
scale that it is more convenient to consider its time average
$\overline{C}$, defined as the integral of $C$ over one period
of oscillation divided by the period. The dynamic heat capacity
$\overline{C}(\omega)$ presents a dependence on
the angular frequency $\omega$ of temperature oscillations,
that is, a dispersion on $\omega$, 
which allows the introduction of a complex heat capacity $C_c$
\cite{gobrecht1971,birge1985,jeong1997,schawe1997,
hohne1997,simon1997,jones1997,baur1998,claudy2000,garden2007c}.
According to Birge and Nagel \cite{birge1985}, the dispersion
of the real part of the complex heat capacity requires
by the Kramers-Kronig relations the existence of an imaginary part.
The dynamic heat capacity $\overline{C}$
is the real part of the complex heat capacity, $\overline{C}={\rm Re}(C_c)$.
The question then arises as to the meaning of the imaginary part
of the complex heat capacity \cite{birge1985,jeong1997,schawe1997,
hohne1997,simon1997,jones1997,baur1998,claudy2000,garden2007c}.

Usually, the imaginary part ${\rm Im}(C_c)$ of the complex heat capacity
is related to the absorption of heat
by the sample, but in modulation temperature experiments there is
no net absorption of energy because in a cycle the heat absorbed equals
the heat released. In spite of this fact, there is a net increase in entropy
in a cycle, which induces an entropy flux to outside.
The imaginary part of the complex
heat capacity is interpreted as either the net entropy flux during one cycle, 
or the entropy produced in a cycle $S_{\rm cy}$ 
\cite{birge1985,jeong1997,schawe1997,hohne1997,simon1997,
jones1997,baur1998,claudy2000,garden2007c}.
These two interpretations are not independent from each other because
the net entropy flux and net entropy production 
are equal in the dynamic steady state.
Birge and Nagel \cite{birge1985} argue that
\beq
S_{\rm cy} = \pi\varepsilon^2\,{\rm Im}(C_c),
\label{17}
\eeq
where $\varepsilon=T_1/T_0$ is the ratio between the amplitude $T_1$
of temperature oscillations and the mean temperature $T_0$.
An interpretation given by Hohne relates 
the imaginary part to the area of the cycle obtained by plotting
the heat flux versus temperature \cite{hohne1997},
\beq
A_{\rm cy} = \pi \,T_1^2 \,{\rm Im}(C_c).
\eeq


Any system in out of equilibrium will be in a state of continuous
production of entropy. In the non-equilibrium stationary state,
which is approached for large times, the rate of entropy
production $\Pi$ becomes equal to the  entropy flux $\Phi$.
In the present case of temperature modulation, the
system does not properly reach a stationary state but
reaches, after a transient regime, a dynamic
stationary state in which the various quantities oscillates
with the same frequency. In this case the time average of the
entropy flux $\overline{\Phi}$ becomes equal to the time
average of the rate of entropy production $\overline{\Pi}$.
This quantity is never negative and vanishes in the thermodynamic
equilibrium, that is, when the $\omega\to0$.
 
Here we analyze the behavior of temperature modulated systems in the light
of the stochastic thermodynamics of systems with continuous phase space
\cite{tome2006,tome2010,broeck2010,spinney2012,zhang2012a,
seifert2012,santillan2013,luposchainsky2013,wu2014,tome2015,tome2015book}.
More precisely, we consider the approach based on the 
Fokker-Planck-Kramers equation \cite{tome2010,tome2015}.
From the present approach we calculate exactly the heat capacity,
of a harmonic solid and show the following general results.
For small amplitudes of the temperature oscillations,
the heat flux behaves sinusoidally, a behavior
that permits the definition of complex heat capacity.
The imaginary part of the complex heat capacity ${\rm Im}(C_c)$ is proportional
to the rate of production of entropy, or more precisely,
to the time average $\overline{\Pi}$ of this quantity.
The poles of the complex heat capacity $C_c$ are 
equal to the eigenvalues of the unperturbed evolution operator
multiplied by the imaginary unit. In addition, the poles
are located in the lower half plane of the complex frequency.

\section{Non-equilibrium stochastic thermodynamics}

The basic equation of stochastic thermodynamics that 
we use here is the Fokker-Planck-Kramers (FPK) equation
describing the time evolution of a system of interacting
particles at a given temperature $T$ \cite{tome2010,tome2015}. 
The FPK equation governs the time evolution of the probability
$P(x,v,t)$ of state $(x,v)$ at time $t$, where $x$ and $v$
denotes the collection of positions and velocities of particles,
and is given by \cite{tome2010,tome2015}
\beq
\frac{\partial P}{\partial t} = - \sum_i
\left( v_i\frac{\partial P}{\partial x_i}
+\frac1{m}f_i \frac{\partial P}{\partial v_i}
+ \frac{\partial J_i}{\partial v_i}\right),
\label{1}
\eeq
where
\beq
J_i = - \gamma v_i P - \frac{\gamma \kB T}{m}
\frac{\partial P}{\partial v_i},
\eeq
and $m$ is the mass of each particle, $\gamma$ is the
dissipation constant, and $f_i=-\partial V/\partial x_i$
is the force acting on particle $i$.
If $T$ is kept constant, the probability distribution approaches
for long times the Gibbs equilibrium distribution associated
to the energy function
\beq
E = \sum_i \frac12 m v_i^2 + V.
\eeq
When $T$ is time dependent, the Gibbs distribution is no longer
a solution of the FPK equation and we will resort to the method
explained below.

The time variation of the average of any state function 
is obtained by multiplying the right and left-hand side
of the FPK equation and integrating over all positions and
velocities. Accordingly, the time variation of the mean energy 
$U=\la E\ra$ is found to be
\beq
\frac{dU}{dt} = -\Phi_{\rm q}, 
\label{7}
\eeq
where $\Phi_{\rm q}$ is the heat flux {\it from} the system {\it to}
the environment,
\beq
\Phi_{\rm q} = n\gamma(m\la v_i^2\ra - \kB T).
\eeq
The first term on the right-hand site is
understood as the heating power and the second as the heat loss.
We are considering that $\la v_i^2\ra$ is independent of $i$,
and $n$ is the number of degrees of freedom. 

The entropy is assumed to have the Gibbs form
\beq
S = - \kB \int P \ln P dxdv.
\eeq
Its time variation can be split into two parts,
\beq
\frac{dS}{dt} = \Pi - \Phi,
\eeq
where $\Pi$ is the rate of entropy production \cite{tome2010,tome2015},
\beq
\Pi = \frac{nm}{\gamma T} \int \frac{J_i^2}{P}dxdv,
\label{11}
\eeq
and $\Phi$ is the entropy flux {\it from} the system
{\it to} the environment,
\beq
\Phi = \frac{\Phi_{\rm q}}{T}.
\label{43}
\eeq
Notice that the rate of entropy production $\Pi$
is always positive, as follows from the right-hand side
of equation (\ref{11}), vanishing only in equilibrium
when $J_i=0$. The dynamic heat capacity is defined by
\beq
C = \frac{-\Phi_{\rm q}}{dT/dt}.
\label{41}
\eeq
and equals $C=(dU/dt)/(dT/dt)$. 
It should be remarked that the dynamic heat capacity
does not share the property $C_0\geq0$ of the static
heat capacity, and it may be negative. This happens
when the variation in temperature is so fast that
the flux of heat is in opposition of phase.
The positivity of heat capacity, which is the hallmark
of thermodynamic stability of the system, is required
if the system is in or near equilibrium, which is not
the case here, in  general.

\section{Complex heat capacity}

Next we consider that the temperature oscillates in time around $T_0$
with an angular frequency $\omega$ and amplitude $T_1$,
according to
\beq
T = T_0 + T_1\cos\omega t.
\label{1a}
\eeq
To solve the FPK equation for this case we start by defining
the differential operators ${\cal F}$ and ${\cal K}$ by
\beq
{\cal F}P = - \sum_i \left(
v_i\frac{\partial P}{\partial x_i}
+\frac1{m}f_i \frac{\partial P}{\partial v_i}
- \gamma\frac{\partial v_i P}{\partial v_i}\right) + {\cal K}P,
\label{1b}
\eeq
and
\beq
{\cal K}P =  \frac{\gamma\kB T_0}{m} \sum_i \frac{\partial^2 P}{\partial v_i^2}.
\eeq
The FPK equation (\ref{1}) takes the form
\beq
\frac{\partial P}{\partial t} = {\cal F}P + \varepsilon \cos\omega t {\cal K}P,
\label{2}
\eeq
where $\varepsilon=T_1/T_0$. 
When $\varepsilon$ is zero, and for long times, the probability
distribution $P$ approaches the Gibbs distribution
$P_0 = \exp\{-E/\kB T_0\}/Z$, which gives ${\cal F}P_0=0$.
To find a solution which is correct up to first order in $\varepsilon$, it suffices
to replace $P$ in the last term of equation (\ref{2}) by the 
time independent equilibrium probability distribution $P_0$,
\beq
\frac{\partial P}{\partial t} = {\cal F}P + \varepsilon \cos\omega t {\cal K}P_0.
\label{2a}
\eeq

To solve equation (\ref{2a}), we start by noticing 
that its solution can be seen as the real part part of a complex quantity, 
that is, $P={\rm Re}(P_c)$, and that $P_c$ obeys the equation
\beq
\frac{\partial P_c}{\partial t} = {\cal F}P_c + \varepsilon e^{-i\omega t} {\cal K}P_0.
\label{2b}
\eeq
Formally, $P_c$ is the solution of the FPK equation when $T$ is replaced by
\beq
T_c=T_0+T_1 e^{-i\omega t}.
\label{19}
\eeq
The solution of equation (\ref{2b}) is of the type
\beq
P_c = P_0 + \varepsilon Q e^{-i \omega t},
\label{6}
\eeq
where $Q$ is a time independent complex quantity,
obeying the equation
\beq
-i\omega Q = {\cal F}Q + {\cal K}P_0,
\label{13}
\eeq
where we have taken into account that ${\cal F}P_0=0$.
The problem of solving equation (\ref{2b}), and therefore
(\ref{2a}), is reduced to solving equation (\ref{13}). 

If equation (\ref{13}) is solved, the average $\la g\ra_c$
of a state function $g$ with respect to $P_c$ is
\beq
\la g\ra_c = \la g\ra_0 + \varepsilon\la g\ra_1 e^{-i \omega t},
\eeq
where the subscripts 0 and 1 denote the averages with
respect to $P_0$ and $Q$, respectively, 
and the average with respect to $P$ will be $\la g\ra = {\rm Re}\la g\ra_c$.
Accordingly, the heat flux $\Phi_{\rm q}={\rm Re}(\Phi_{\rm q}^c)$, where
\beq
\Phi_{\rm q}^c = n\gamma(m\la v_i^2\ra_c - \kB T_c)
= \phi \varepsilon e^{-i\omega t},
\eeq
and we are using the abbreviation
\beq
\phi = n\gamma(m\la v_i^2\ra_1 - \kB T_0),
\label{47}
\eeq
and the result $\la v_i^2\ra_0=\kB T_0$.

The complex heat capacity $C_c$ is defined in a way analogous to
the definition of the dynamic heat capacity (\ref{41}),
\beq
C_c = \frac{-\Phi_{\rm q}^c}{dT_c/dt}
= \frac{\Phi_{\rm q}^c}{i\omega T_1 e^{-i\omega t}} = \frac{\phi}{i\omega T_0},
\label{42}
\eeq
and is time independent.
From the definition (\ref{41}) of the dynamic heat capacity,
\beq
C = \frac{{\rm Re}(\Phi_{\rm q}^c)}{\omega T_1\sin\omega t} 
= \frac{1}{\omega T_0}{\rm Re} \frac{\phi e^{-i\omega t}}{\sin\omega t},
\eeq
and from the relation (\ref{43}) between entropy flux $\Phi$
and heat flux, the entropy flux is
\beq
\Phi = \frac{{\rm Re}(\Phi_{\rm q}^c)}{T_0 + T_1\cos\omega t} 
= \frac{1}{T_0}{\rm Re}\frac{\phi e^{-i\omega t}}{r + \cos\omega t},
\eeq
where $r=T_0/T_1>1$.
To determine the time averages of $C$ and $\Phi$ we use the
following results
\beq
\frac{1}{2\pi}\int_0^{2\pi}\frac{e^{-i\theta}}{\sin\theta}d\theta = \frac{1}{i},
\eeq
\beq
\frac{1}{2\pi}\int_0^{2\pi}\frac{e^{-i\theta}}{r + \cos\theta}d\theta = -\lambda,
\eeq
where $\lambda=(r/\sqrt{r^2-1})-1$, which are easily obtained by the
methods of residues. Using these results we find
\beq
\overline{C} = \frac{1}{\omega T_0}{\rm Re}(\phi/i) = {\rm Re}(C_c),
\label{50a}
\eeq
\beq
\overline{\Pi}=\overline{\Phi} = -\frac{\lambda}{T_0}{\rm Re}(\phi)
= \lambda\omega {\rm Im}(C_c),
\label{50b}
\eeq
where we have taken into account that in the dynamic non-equilibrium stationary state
the time averages of the flux of entropy and rate of entropy production are equal,
$\overline{\Pi}=\overline{\Phi}$. 
The two results (\ref{50a}) and (\ref{50b}) are summed up as
\beq
C_c = \overline{C} + \frac{i}{\lambda\omega}\overline{\Pi}.
\label{50}
\eeq

For small values of $\varepsilon$, we find $\lambda=\varepsilon^2/2$ and
\beq
S_{\rm cy} = \frac{2\pi}{\omega}\overline{\Pi}
= \pi \varepsilon^2 \,{\rm Im}(C_c).
\label{8}
\eeq
which is the result (\ref{17}) of Birge and Nagel \cite{birge1985},
connecting the production of entropy in a cycle and the imaginary part of $C_c$.
The area of the cycle in the temperature versus heat flux plane is
\beq
A_{\rm cy} = \frac{2\pi}{\omega}\overline{T\Phi_{\rm q}} 
= \pi \,T_1^2 \,{\rm Im}(C_c)
\eeq 
which is the result of Hohne \cite{hohne1997},
connecting $A_{\rm cy}$ to the imaginary part of $C_c$.


We now proceed to an analysis of the poles of $C_c$. 
We prove that they are located in the
lower half plane of the complex frequency assuring that the
Kramers-Kronig relations \cite{landau1958,groot1962}
connect the real and imaginary part of $C_c$.
Let us denote by $\psi_\ell$ and $\Lambda_\ell$ the eigenfunctions and
eigenvalues of ${\cal F}$. Notice that
the probability distribution $P_0=\psi_0$ is an eigenfunction with
zero eigenvalue. Expanding $Q$ in eigenfunctions of ${\cal F}$,
\beq
Q = \sum_{\ell} q_\ell \psi_\ell,
\eeq
and replacing into equation (\ref{13}) for $Q$, we find
\beq
q_\ell = - \frac{\la{\cal K} \chi_\ell\ra_0 }{i\omega + \Lambda_\ell},
\label{46}
\eeq
where $\chi_\ell$ are the eigenfunctions of the adjoint operator
${\cal F}^\dagger$ of the operator ${\cal F}$.
The coefficient $q_0=0$ because $\psi_0^\dagger=1$.
Since $C_c$ is an average with respect to $Q$, the poles of $C_c$
are the possible poles of $q_\ell$,  $\ell\neq0$, which are $\omega_\ell = i\Lambda_\ell$
as can be inferred from (\ref{46}), and we see that they are related to the 
nonzero eigenvalues $\Lambda_\ell$ of the unperturbed ${\cal F}$ operator. 
Considering that the nonzero eigenvalues of the 
operator ${\cal F}$ are complex number with negative real part,
$\Lambda_\ell=-\Lambda_\ell^0 + i \Lambda_\ell^1$, we find
\beq
\omega_\ell = - \Lambda_\ell^1 - i\Lambda_\ell^0,
\eeq
and the poles are located in the lower half plane of complex $\omega$
because $\Lambda_\ell^0>0$.

Looking at expression (\ref{42}), it seems at first sight that
$\omega=0$ is a pole. But that is not the case.
To be a pole, $C_c$ should diverge in the limit $\omega\to0$,
but in fact it is finite and equal to the static heat capacity.
The absence of poles in the upper half plane guarantees that $C_c$
is analytic in this region and that the 
real and imaginary part of $C_c$ obey the Kramers-Kronig relations.

\section{Harmonic solid}

In the following we carry out an exact calculations to
determine $\overline{C}$ and $\overline{\Pi}$
for a harmonic solid whose energy is given by 
\beq
E=\frac12 \sum_i mv_i^2 + \frac12 \sum_{ij} K_{ij} x_ix_j,
\eeq
where $K_{ij}$ are the elements of a $n\times n$ real symmetric matrix $K$
with nonnegative eigenvalues.
According to our approach developed above it suffices to
find the real and complex part of $C_c$, which is obtained
by calculating $\la v_i^2\ra_1$ and $\phi$. 
Let us use the notation $B_{ij}=\la x_ix_j\ra_1$, $A_{ij}=\la v_iv_j\ra_1$,
and $C_{ij}=\la x_iv_j\ra_1$. 
From equation (\ref{13}) the averages with respect to $Q$ can be
calculated by
\beq
-i\omega \la g\ra_1 = \la {\cal F}^\dagger g\ra_1 + \la {\cal K} g\ra_0, 
\label{21}
\eeq
a relation valid for any state function $g$.
From equation (\ref{21}), we find a closed set 
of equations for the matrices $A$, $B$ and $C$, whose entries are
$A_{ij}$, $B_{ij}$ and $C_{ij}$, respectively,
\begin{eqnarray}
(-i\omega + 2\gamma )A
&=& - \frac1m(K C + C^{\sf T} K) +\frac{2\gamma\kB T_0}{m}I,
\label{51a} \\
-i\omega B &=& C + C^{\sf T},
\label{51b} \\
(-i\omega +\gamma) C &=& A - \frac1m B K.
\label{51c}
\end{eqnarray}

\begin{figure}
\centering
\epsfig{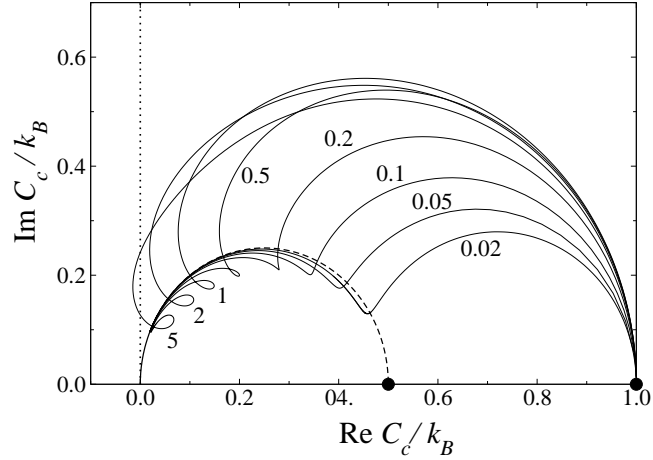}
\caption{Imaginary versus real part of the complex
heat capacity (\ref{40}) for $\omega_\ell^2/\gamma^2=0$ 
(dashed line) and other values indicated (solid lines).
The thermodynamic equilibrium, $\omega=0$, is indicated by a 
full circle. The vertical dotted line indicates the zero value
of the real part.}
\label{imre}
\end{figure}

To solve these equations we use a unitary transformation that diagonalizes
the matrix $K$, transforming it in a diagonal matrix consisting of the
eigenvalues $m\omega_\ell^2$ of $K$. By inspection of the equations
it is seen that the matrices
$A$, $B$ and $C$ become also diagonal. Denoting the diagonal elements
denoted by $a_\ell$, $b_\ell$ and $c_\ell$, the equations for these
variables are
\begin{eqnarray}
(-i\omega + 2\gamma )a_\ell &
=& - 2\omega_\ell^2 c_\ell + \frac{2\gamma\kB T_0}{m}, \\
-i\omega b_\ell &=& 2 c_\ell, \\
(-i\omega +\gamma) c_\ell &=& a_\ell - \omega_\ell^2 b_\ell.
\end{eqnarray}

The quantity $\phi$, given by (\ref{47}), and connected to the complex
heat capacity by $C_c=\phi/i\omega T_0$, becomes 
$\phi=\gamma {\rm Tr}(A-\kB T_0 I)$. Considering
that the trace is invariant under a unitary transformation we find
$\phi = \sum_\ell \phi_\ell$, where $\phi_\ell=\gamma(a_\ell-\kB T_0)$.
Solving the above equations for $a_\ell$, $b_\ell$ and $c_\ell$, 
we determine $\phi_\ell$ and 
\beq
\phi = \kB T_0 \sum_\ell \frac{i\omega\gamma}{\gamma - i\omega}\,
\frac{\omega^2-4\omega_\ell^2+i\omega\gamma}{\omega^2 - 4\omega_\ell^2 + 2i\omega\gamma},
\eeq
and reach the following result for the complex heat capacity
\beq
C_c = \kB \sum_\ell \frac{\gamma(\omega^2-4\omega_\ell^2+i\omega\gamma)}
{(\gamma - i\omega)(\omega^2 - 4\omega_\ell^2 + 2i\omega\gamma)}.
\label{40}
\eeq
The imaginary part versus the real part of $C_c$ is shown in figure
\ref{imre}. Notice that the real part, which is identified as the
dynamical heat capacity, may be negative as seen in one of the
curves of this figure.

\begin{figure}
\centering
\epsfig{file=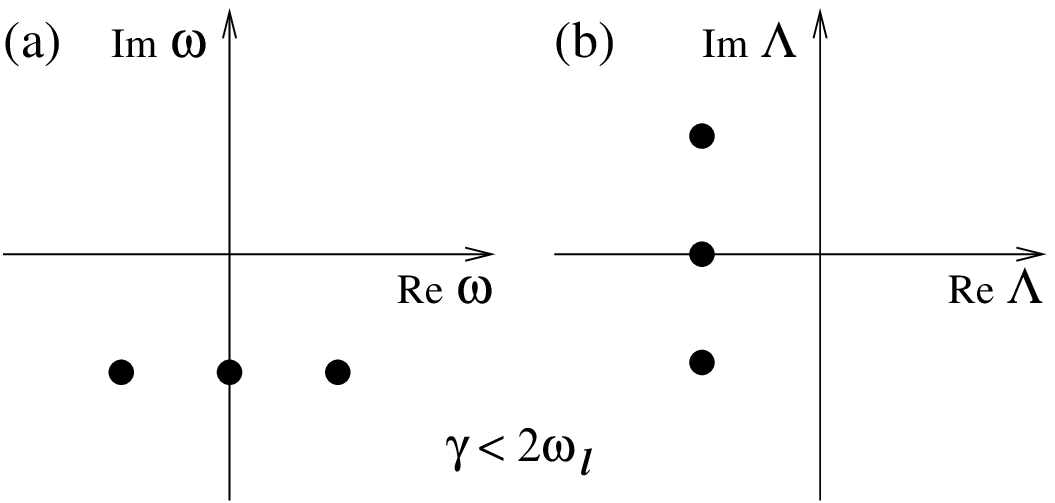,width=7.5cm}
\hfill
\epsfig{file=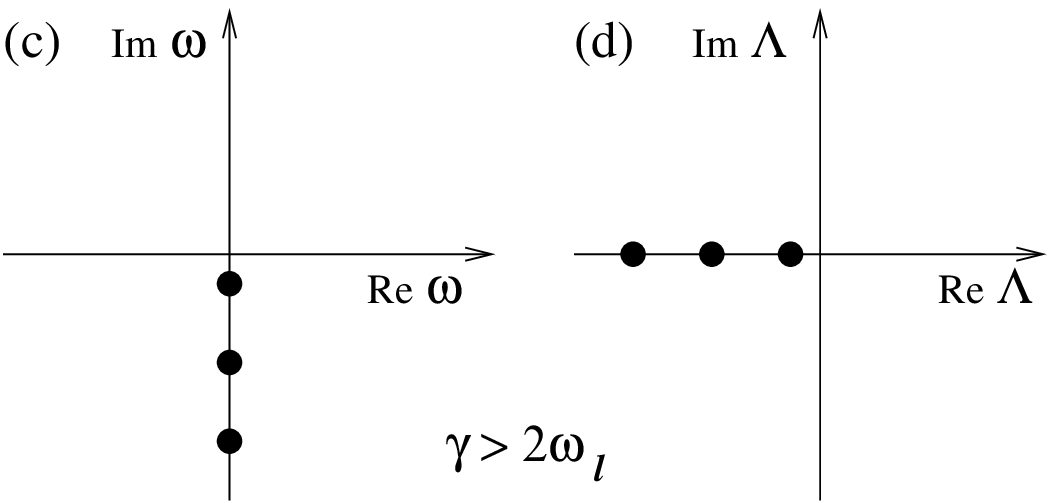,width=7.5cm}
\caption{(a) and (c) show the poles $\omega$ of the complex heat
capacity $C_c$; and (b) and (d) show the eigenvalues $\Lambda$ of
the unperturbed evolution operator ${\cal F}$. They are
related by $\omega=i\Lambda$, that is, the poles are obtained
from the eigenvalues through an anticlockwise rotation of $\pi/2$.}
\label{poles}
\end{figure}

Each term of the complex heat capacity $C_c$ has three poles, all
located in the lower half-plane of the complex $\omega$, 
as shown in figure \ref{poles}.
If $4\omega_\ell^2 > \gamma^2$, the poles are
\beq
\omega = \left\{
\begin{array}{l}
-i\gamma \\
-i\gamma \pm\sqrt{4\omega_\ell^2-\gamma^2}
\end{array}
\right.
\eeq
If $4\omega_\ell^2 < \gamma^2$, the poles are pure imaginary and given by
\beq
\omega = \left\{
\begin{array}{l}
-i\gamma \\
-i\gamma \pm i \sqrt{\gamma^2-4\omega_\ell^2}
\end{array}
\right.
\eeq
The complex heat capacity (\ref{40}) is thus analytic in the upper
half-plane and vanishes as $1/|\omega|$ so that it obeys theKramers-Kronig relations.

The real part of $C_c$ gives the time average of the dynamic heat capacity,
\beq
\overline{C} = \kB \sum_\ell \frac
{2\gamma^2(\omega^4-6\omega_\ell^2\omega^2 + 8\omega_\ell^4 +\gamma^2\omega^2)}
{(\omega^2+\gamma^2)(\omega^4-8\omega_\ell^2\omega^2 + 16\omega_\ell^4 +4\gamma^2\omega^2)},
\eeq
and its imaginary part
gives the time average of the rate of entropy production,
\beq
\overline{\Pi} = \kB \sum_\ell\frac
{\lambda\gamma\omega^2(\omega^4-8\omega_\ell^2\omega^2
+ 16\omega_\ell^4 + 4\omega_\ell^2\gamma^2+\gamma^2\omega^2)}
{(\omega^2+\gamma^2)(\omega^4-8\omega_\ell^2\omega^2 + 16\omega_\ell^4 +4\gamma^2\omega^2)}.
\eeq
The rate of production of entropy $\overline{\Pi}$ is positive as demanded by 
expression (\ref{11}). It vanishes when $\omega=0$ and reaches the value 
$\lambda\gamma n\kB$ when $\omega\to\infty$. The dynamic heat capacity
vanishes in this limit. When $\omega=0$ it yields the value $n\kB$, which is the
classical heat capacity of a harmonic solid with $n$ degrees of freedom.

The eigenvalues of the operator ${\cal F}$ is obtained
by solving the eigenvalue
equation ${\cal F}\psi=\Lambda\psi$. They can be found by
assume a solution of the type
$\psi=P_0G$ where $G(x,v)$ is the quadratic form
\beq
G = G_0+\frac12\sum_{ij}(A_{ij}v_iv_j + B_{ij}x_ix_j + C_{ij}x_iv_i).
\eeq
The replacement of this form into the eigenvalue equation,
yields a set of equations for $A$, $B$ and $C$ which is found to be that given by
equations (\ref{51a}), (\ref{51b}), and (\ref{51c}), with $-i\omega$ replaced
by $\lambda$ and without the last term of (\ref{51a}). Proceeding in a
similar manner, we find the eigenvalues of ${\cal F}$, which are
shown in figure \ref{poles}.
If $4\omega_\ell^2 > \gamma^2$, the eigenvalues are
\beq
\Lambda = \left\{
\begin{array}{l}
-\gamma \\
-\gamma \mp i\sqrt{4\omega_\ell^2-\gamma^2}
\end{array}
\right.
\eeq
If $4\omega_\ell^2 < \gamma^2$, the eigenvaues are real and given by
\beq
\Lambda = \left\{
\begin{array}{l}
-\gamma \\
-\gamma \pm \sqrt{\gamma^2-4\omega_\ell^2}
\end{array}
\right.
\eeq
This confirms the
result that the poles of the complex heat capacity are indeed given by
$\omega=i\Lambda$.

\section{Conclusion}

\vskip-0.2cm

Systems subject to a sinusoidal time dependent temperature 
are in a state of non-equilibrium and permanent production of entropy.
In such a situation it is
appropriate to analyzed them under the light of stochastic non-equilibrium
thermodynamics. In theoretical studies of thermal modulation, the
sinusoidal behavior of the heat flux is usually taken from
granted. Using stochastic thermodynamics, we have demonstrated that this assumption
is generally valid as long as the amplitude of the oscillations of temperature
is small. That is, the heat flux behaves sinusoidally in time as the temperature
but with a phase shift. If the amplitude
is not small, the heat flux is still a periodic function of time
but is not pure sinusoidal. Anyhow, the smallness
of the amplitude of the temperature modulation is always met in
experimental investigations so that we do not need to go beyond the linear
approximation. In the case of harmonic forces, on the other hand,
we have shown that it is not necessary to assume that the amplitude is small.
In this case the heat flux will be always sinusoidal
if the temperature is sinusoidal. 

The pure sinusoidal behavior of the heat flux allowed
the definition of the complex heat capacity whose
real part is the dynamical heat capacity.
The imaginary part of the complex heat capacity 
is interpreted as related to the net entropy flux in a cycle  
which equals the net production of entropy in a cycle. Here
we have demonstrated this significant result achieved by
Birge and Nagel \cite{birge1985} in their experimental
investigations on the specific-heat spectroscopy. Our procedure 
is distinct from other approaches and is accurate enough
to reach a precise relation between the imaginary part
of the complex heat capacity and the entropy production,
including the correct prefactor. 

We have found and demonstrated a connection between the poles
of the complex heat capacity and the eigenvalues of the unperturbed
evolution operator. The former being the latter multiplied by the imaginary unit.
In addition, we have shown that
all poles are located in the lower half-plane of the complex frequency
plane. The absence of 
poles in the upper half-plane implies the analyticity of the 
complex heat capacity in this region
which in turn leads to the Kramers-Kronig relations. 
These results constitute some of the main results of the present study.

When the forces are harmonic, it is possible to solve exactly the
FPK equation for any amplitude of the sinusoidal temperature.
That is, it is possible to solve the FPK equation for any value of
$\varepsilon$, not necessarily small. This is possible because in this
case the time dependent probability distribution is Gaussian so that
the only correlations that are needed are the pair correlations.
We have carried out an exact calculations of the pair correlations
from which we find exactly the complex heat capacity
as a function of the frequency and obtained
the frequency dispersion of the dynamic heat
capacity and of the rate of entropy production.

It should be mentioned finally that all results
obtained here, which include in the first place
the possibility of the definition of a complex heat capacity,
are founded on the sinusoidal behavior of the flux of heat.
As we have demonstrated, this happens when the amplitude
of the time oscillations of the temperature is small. It also
happens when forces are harmonic, in which case the amplitude of temperature
oscillations do not have to be small.


\end{document}